\documentclass[3p]{elsarticle}
\usepackage{amssymb,amsmath}

\usepackage{graphicx}
\usepackage{cancel}
\usepackage[ruled]{algorithm2e}
\usepackage{hyperref}
\usepackage{multirow}

\usepackage[colorinlistoftodos,textwidth=4cm,shadow]{todonotes}
\newcounter{Denis}

\newcounter{Igor}

\newcounter{George}

\date{}

\title {\textbf{
		A parameteric class of composites with a large achievable range of effective elastic properties 
		}}

\author[add1]{Igor Ostanin \footnote{Corresponding author, tel: +79150174677, e-mail:i.ostanin@skoltech.ru}
}
\author[add1]{George Ovchinnikov}

\author[add2]{Davi Colli Tozoni}

\author[add1,add2]{Denis Zorin}

\address[add1]{Skolkovo Institute of Science and Technology, Nobel St. 3, Moscow, 143026, Russia}
\address[add2]{New York University, 719 Broadway, New York, 10003, USA}

\begin{document}

\begin{abstract}
In this paper, we study an instance of the  \emph{G-closure} problem for two-dimensional
periodic metamaterials. Specifically, we consider composites with isotropic homogenized elasticity tensor,
obtained as a mixture of two isotropic materials. We focus on the case when one material is has zero stiffness i.e., single-material structures with voids. This problem is important, in particular, in the context of designing small-scale structures for metamaterials that can be manufactured using additive fabrication. 
A range of effective metamaterial properties can be obtained this way using a single base material. 

We demonstrate that two closely related simple parametric families based on the structure proposed by
O. Sigmund in \cite{Sigmund2000} attain good coverage of the space of isotropic properties satisfying
Hashin-Shtrikman bounds. In particular, for positive Poisson's ratio, we demonstrate that Hashin-Shtrikman
bound can be approximated arbitrarily well, within limits imposed by numerical approximation: a strong
evidence that these bounds are achievable in this case.  For negative Poisson's ratios, we numerically
obtain a bound which we hypothesize to be close to optimal, at least for metamaterials with rotational
symmetries of a regular triangle tiling.

\end{abstract}

\maketitle

\section{Introduction}
\label{sec:intro}

The relationship between  the geometric structure of a periodic composite material and its effective properties (elastic properties in particular) is a central question in mechanics of composites. 
The direct problem -- finding the effective elastic properties from the known periodic structure of the composite material -- is solved by means of homogenization theory. The inverse problem (finding a periodic structure yielding specific material properties) is more challenging: in general, it requires solving a non-convex non-linear PDE constrained optimization problem with highly nonunique solution (inverse homogenization).  Most importantly, it is not known in which cases this problem has feasible solutions, \textit{i.e.}, for which target elasticity tensors a corresponding composite structure with a given base material exists.
This problem is known as the G-closure problem \cite{Allaire2002}.  While for the heat conduction  explicit solutions
are known in some cases, for elasticity this is a long-standing problem. The outer bounds of the feasible region
in the space of elasticity tensors are given by the well-known Hashin-Shtrikman bounds \cite{Hashin} on the bulk and shear moduli, as well as their refinement \cite{CGbounds}.  For positive Poisson ratios, the answer is provided in the recent work \cite{Milton2017}. The right Hashin-Shtrikman bounds are achieved by sequential laminates, \textit{i.e.}, assuming infinite geometric complexity and separation of scales (\textit{i.e.}, very large differences in periods of  laminations in sequence). In general, these materials cannot be practically realized, although one can manufacture, with considerable difficulty, finite-resolution approximations.

We explore solutions to this problem numerically, using specific simple parametric families of structures inspired by several previous papers. We show that with modest geometric complexity, and no separation of scales, one can approach the boundaries closely. Thus, a large fraction of our structures can be, in principle, manufactured and can serve as a starting point for further simplification. Our experiments provide an estimate on the topological complexity required to achieve a particular percentage of coverage of the 
area in the space of elastic properties defined by  Hashin-Shtrikman bounds.

\begin{figure}  
	\includegraphics[width=16cm]{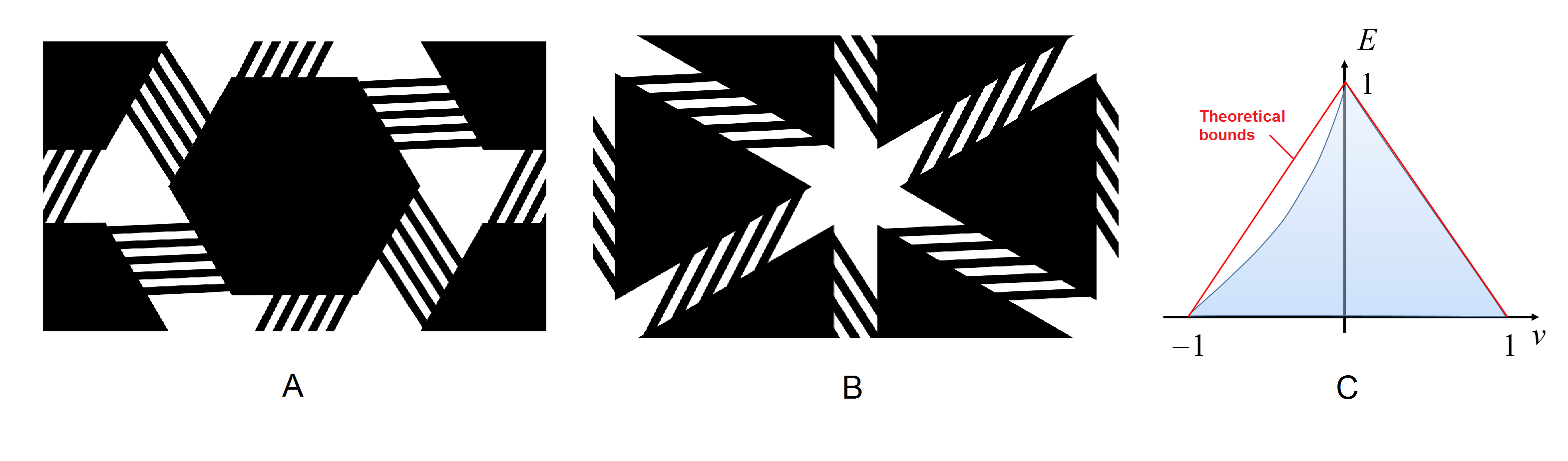}
	\caption{ (A,B) parametric structure families, (C) Domains of achievable effective elastic properties and Hashin-Shtrikman theoretical bounds ( $E_b = 1$, $\nu_b = 0$ )   in $(E, \nu)$ domain.} 
        \label{Schematics}
\end{figure}

\paragraph{Contributions}
We explored two families of structures (depicted in Figure~\ref{Schematics}(A, B)), using  a hexagonal base cell, and symmetric with respect to rotation of the cell by $n\pi /3$, which ensures that their effective elastic tensors are perfectly isotropic. Each structure has only four parameters. 
We observe that these structures have the following properties:

\begin{itemize}

\item For positive effective Poisson's ratio and volume fractions approaching one, the bounds of the Hashin-Shtrikman region are approached arbitrarily closely, within limits of numerical accuracy.  
\item For negative effective Poisson's ratio and volume fractions approaching one, chirality parameter allows us to cover a larger fraction of the region defined by the Hashin-Shtrikman bounds than all previously known.  	
\item The low number of parameters of the structure allows for simple mapping of material parameters to structure parameters, potentially avoiding inverse homogenization entirely. 
\item We show that for volume fractions different from one and positive effective Poisson's ratios, proposed structures are also close to the Hashin-Shtrikman bounds.
\item For negative effective Poisson's ratios, we show evidence that our structures are close to locally optimal (i.e., increasing the number of structure parameters is not likely to lead to improvements) 
\item  Our family of structures in 2D covers a larger fraction of the Hashin-Shtrikman domain than all previously known families.

\end{itemize}

Figure~\ref{Schematics}(C) shows  the set of Young's moduli and Poisson's ratios that could be achieved by tuning the parameters of structures (A) and (B), as related to the known theoretical bounds on these moduli (see the discussion in the next section).

\paragraph{Remark} To illustrate material property coverage, in most of our figures we use Young's modulus -- Poisson's ratio coordinates, vs. bulk-shear modulus more commonly found in the literature: the distance in this space captures the differences in material behavior in a more intuitive way.  We discuss the relationship between these
two parameterizations of material properties below. 
   
\section{Background and related work}
\label{sec:related}

\paragraph{Bounds on elastic properties of isotropic materials}
The elastic tensor characterizing properties of a periodic composite material  is  obtained by homogenization over a base cell of the composite structure (Section~\ref{sec:methods}). In this work, we consider composite structures that are isotropic due to spatial symmetries of the base cell. Such composites can be described by two independent elastic moduli (bulk and shear moduli or Young's modulus and Poisson's ratio). 
The best known bounds for composite properties are Hashin-Shtrikman bounds \cite{Hashin}. In case of well ordered strong ($\kappa_s, \mu_s$) and weak ($\kappa_w, \mu_w$) phases (  $(\kappa_s^{HS} - \kappa_w^{HS}) (\mu_s^{HS} - \mu_w^{HS}) >0$), the HS bounds are:

\begin{align*}
\kappa_{l}^{HS}& < \kappa < \kappa_{u}^{HS},\quad 
\mu_{l}^{HS} < \mu < \mu_{u}^{HS}, \\
\kappa_{u}^{HS} &=\kappa_s+\frac{1-\phi}{\frac{1}{\kappa_w - \kappa_s}+\frac{\phi}{\kappa_s+\mu_s}},\quad
&\kappa_{l}^{HS} &=\kappa_w+\frac{\phi}{\frac{1}{\kappa_s - \kappa_w}+\frac{(1-\phi)}{\kappa_w+\mu_w}}, \\
\mu_{u}^{HS}    &=\mu_s+\frac{1-\phi}{\frac{1}{\mu_w - \mu_s}+\frac{\phi (\kappa_s + 2 \mu_s) }{2 \mu_s (\kappa_s+\mu_s)}}, 
&\mu_{l}^{HS}    &=\mu_w+\frac{\phi}{\frac{1}{\mu_s - \mu_w}+\frac{(1-\phi) (\kappa_w + 2 \mu_w) }{2 \mu_w(\kappa_w+\mu_w)}}.\hfill
\label {HS4}
\end{align*}

\begin{figure}  
	\includegraphics[width=16cm]{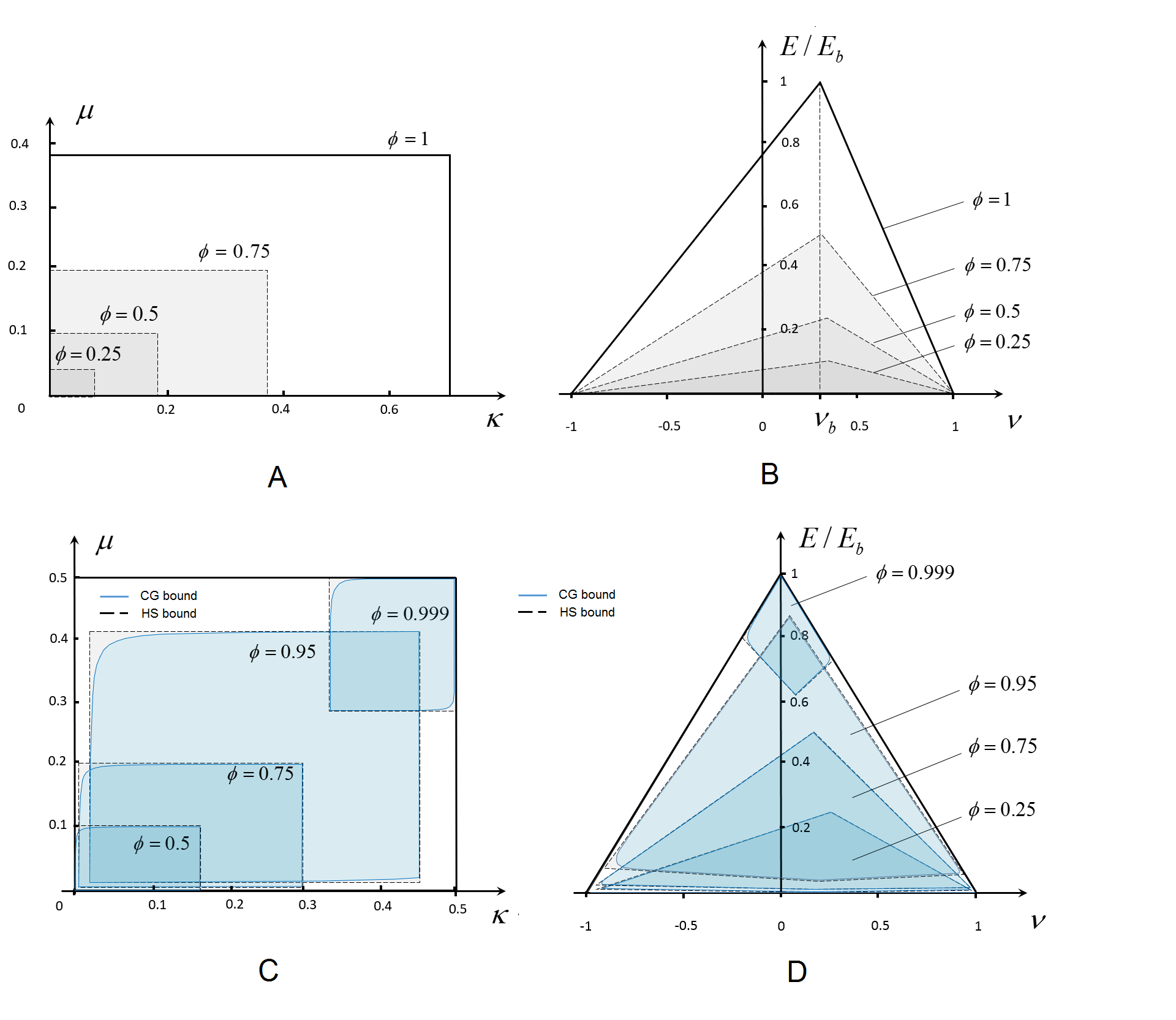}
	\caption{ (A, B) Hashin-Shtrikman bounds for the elastic moduli of isotropic void-material composites: (A) in bulk and shear modulus coordinates and (B) the same bounds transformed to Poisson's ratio and Young's modulus coordinates. (C,D)  Hashin-Shtrikman (grey) and Cherkaev-Gibiansky (blue) bounds for elastic moduli of isotropic bimaterial composite ($\nu_s = \nu_w = 0$, $E_w = 0.001 E_s$) (C) in terms of bulk and shear moduli and (D) in terms of Poisson's ratio and Young's modulus. For brevity, here and below Hashin-Shtrikman bounds are labeled on images as HS bounds and Cherkaev-Gibiansky bounds -- as CG bounds.}
\label{HSFIG}
\end{figure}

Figure~\ref{HSFIG}  shows  Hashin-Shtrikman bounds for void-material (A,B) and bimaterial (C,D) composites, in terms of bulk and shear moduli (A,C) and Young's modulus and Poisson's ratio (B,D) for arbitrary volume fractions $0\ldots 1$ and for several fixed volume fractions.
These bounds are known not to be optimal the general case:  Cherkaev-Gibiansky  bounds are substantially tighter \cite{CGbounds}.  However, in the case when the weak phase has zero elastic tensor,  these bounds exactly coincide with the  Hashin-Shtrikman bounds.  In terms of volume fraction $\phi$, and bulk and shear moduli of the base material equal to $\kappa_b$ and $\mu_b$ respectively,  the bounds on effective bulk moduli $\kappa$ and $\mu$ in this case reduce to:

\begin{equation} \label {HS1}
\begin{split}
&0 < \kappa < \kappa_{u}^{HS},\quad
0 < \mu < \mu_{u}^{HS}, \\ 
&\kappa_{u}^{HS}=\kappa_b+\frac{1-\phi}{-\frac{1}{\kappa_b}+\frac{\phi}{\kappa_b+\mu_b}},\quad 
\mu_{u}^{HS}=\mu_b+\frac{1-\phi}{-\frac{1}{\mu_b}+\frac{\phi (\kappa_b + 2 \mu_b) }{2 \mu_b(\kappa_b+\mu_b)}}.
\end{split}
\end{equation}

Expressing bulk and shear moduli as $\kappa = \frac{E}{2(1 - \nu)}$, $\mu = \frac{E}{2(1 + \nu)}$,  one obtains the isotropic bounds in terms of Young's modulus and Poisson's ratio, forming a triangular region:

\begin{equation} \label {HS2}
\begin{split}
&0<E(\nu) < - 2 C_1  \nu + 2C_1,\quad
0<E(\nu) <   2 C_2 \nu + 2C_2, \\ 
&C_{1}=\frac{E_{b}}{2(1-\nu_{b})}+\frac{E_{b}(1-\phi) }{\phi(1-\nu_{b}^{2})-2(1-\nu_{b})}, \\
&C_{2}=\frac{E_{b}}{2(1+\nu_{b})}+\frac{2 E_{b}(1-\phi)}{\phi (3-\nu_{b}) (1+\nu_{b}) -4(1+\nu_{b})}
.
\end{split}
\end{equation}
where $E_b$ and $\nu_b$ are the Young's modulus and Poisson's ratio of the base material. For the case of $\phi=1$, expressions (\ref{HS1},\ref{HS2}) are reduced to 
\begin{equation} \label {HS3}
\begin{split}
&0<\kappa<\kappa_{b},  0<\mu<\mu_{b}, \\
&E(\nu) < -\frac{E_b}{1-\nu_b} \nu + \frac{E_b}{1-\nu_b}, E(\nu) < \frac{E_b}{1+\nu_b}  \nu + \frac{E_b}{1+\nu_b}  
.
\end{split}
\end{equation}

Note that the rational transformation between $(\kappa, \mu)$ and $(E,\nu)$ coordinates is not globally one-to-one: in the extreme case of vanishing Young's modulus,
both $\kappa$ and $\mu$ go to zero, so the lower side of the triangle in $(E,\nu)$ domain collapses to a single point $\kappa = 0$, $\mu = 0$, except points $\nu = \pm 1$.
These two points in $(E,\nu)$ coordinates correspond to the lines $\kappa = 0$ and $\mu = 0$ in $(\kappa, \mu)$ coordinates.
This in part explains why we view $(E,\nu)$ coordinates as more intuitive: for low $E$, there is a substantial difference in measurable behavior
between, \textit{e.g.}, materials with $\nu = -1$, and $\nu = 1$, while both $\kappa$ and $\mu$ are close to zero. 

\begin{figure} 
	\includegraphics[width=16cm]{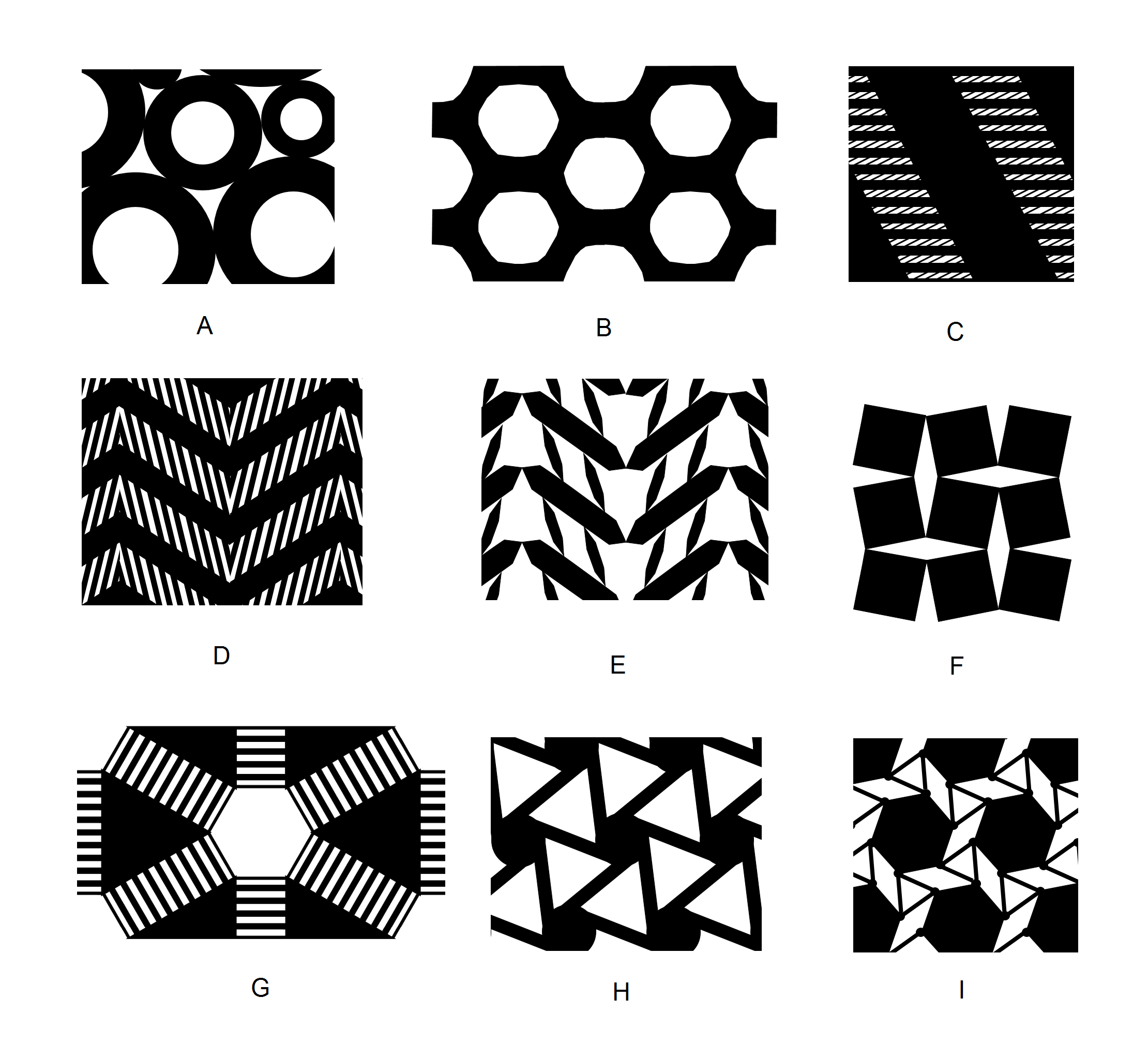}
	\caption{ Known extremal composites (A)  Coated spheres assemblage, exhibiting extremal bulk modulus, (B) Vigdergauz structure, achieving maximum bulk modulus (C) rank 3 sequential laminate, exhibiting maximum bulk and shear modulus  (D) ``Herringbone'' laminate structure providing Poisson's ratio of -1 and (E) and its unimode analog. (F) Auxetic material made of rotating squares \cite{Grima2005} (G) Sigmund extremal structure, providing maximum bulk and minimum shear modulus(H) Chiral isotropic auxetic material\cite{Lakes1996} and (I) similar structure, suggested in \cite{Advanced2011}}
 \label{COMP}
\end{figure}

\paragraph{G-closure} The problem of G-closure was extensively studied. It is known to have a solution for thermal properties; however the problem remains unsolved for elasticity. Related theory is covered in several books \cite{Allaire2002,Torquato2002,Milton2002,Cherkaev2012}.
The first theoretical example of a composite attaining maximum bulk modulus, a random assemblage of coated spheres (Figure~\ref{COMP}(A)) -- was identified by Hashin \cite{Hashin1962}. Another type of structures attaining extremal bulk modulus without separation of scales was found by Vigdergauz for anisotropic \cite{Vigdergauz1989} and later isotropic (Figure~\ref{COMP}(B)) case \cite{Vigdergauz1999}.  Several authors \cite{Lurie1985,Norris1985,Francfort1986,Milton1986}  demonstrated that sequential laminates can achieve extreme bulk and shear modulus simultaneously (Figure~\ref{COMP}(C)). Milton has shown \cite{Milton1992} that sequential laminates can demonstrate negative Poisson's ratio close to -1 (Figure~\ref{COMP}(D)). Milton and Cherkaev \cite{Milton1995} have demonstrated the attainability of G-closure for infinitely rigid and void phases. They describe an approach for constructing composites with any given tensor using elementary structures as building blocks (an example of such structure is given in Figure~\ref{COMP}(E)).  For the same purpose Sigmund adopted laminated regions, and suggested \cite{Sigmund2000} earlier unknown class of extreme isotropic composites (Figure~\ref{COMP}(G)), that display high bulk modulus while maintaining low shear modulus. Our approach is based on this work. 

An important related direction is  the design of materials with negative Poisson's ratio (also known as \textit{auxetic}, or \textit{dilational}). Auxetic materials were first described in the work of Lakes \cite{Lakes1987}, and developed in many papers (see a recent review \cite{Lakes2017} ). Many different types of auxetic materials were identified: beyond already named sequential laminates \cite{Milton1992} and unimode structures \cite{Milton1995} presented in Figures~\ref{COMP}(D,E), there are structures based on rigid rotating units, connected with hinges (Figures~\ref{COMP}(I,F)) \cite{Grima2000,Grima2005,Advanced2011,Solidi2015},  and honeycomb chiral auxetic structures(Figures~\ref{COMP}(H)) \cite{Lakes1996}. We use the latter idea in our current work. 

In their recent work  \cite{Milton2017}, Milton \textit{et al.} identify optimal bounds and corresponding microgeometries for both 2D and 3D anisotropic composites. This work demonstrates that the \emph{right} Hashin-Shtrikmann bound for any volume fraction is achieved by a finite-rank laminate composite. In our work, we demonstrate that the right bound for isotropic 2D composites can be approached arbitrarily closely for any volume fraction by a family of realizable structures, approaching a two-scale composite in the limit, a subclass of  which has been earlier found and studied by Sigmund  \cite{Sigmund2000}.

Topology optimization was used for the purpose of extremal material design (see \cite{Xia2015} for an overview). New periodic structure designs were obtained in \cite{Sigmund1994, Neves2000, Huang2011}.  However, the problem of finding a structure with specific properties is highly nonconvex;  as a result, the optimization often fails to reach the target properties.  Use of filtering techniques  for suppression of checkerboard effects makes it difficult to obtain complex-topology designs which appear to be necessary for target parameters close to the theoretical bounds. Shape optimization is often more successful at achieving specific target material properties, but topology preservation means that initial design topology needs to be obtained by other means.

Wide use of additive fabrication lead to renewed interest in using small-scale structure to achieve specific material behaviors: on the one hand, additive technology makes fabrication of these complex structures possible; on the other hand, these structures allow to manufacture strong parts with lower weight, or objects with continuously variable material properties, \textit{e.g.}, for manufacturing prosthetic devices, or for ``soft'' robotics.  In this context, a number of  additional practical  issues become of importance -- printability of the structure, the absence of extreme stress concentrations that lead to structure damage, its stability towards unwanted nonlinear behavior. Panetta and co-authors \cite{Zorin2015,Zorin2017} presented a framework for the structural design  based on ground state search for topology with subsequent low-parametric shape optimization to achieve the desirable elasticity tensor while satisfying a set of additional constraints (\textit{e.g.} constraint on maximum von Mises stress).

In this work, rather than using topology or shape optimization to solve the problem of G-closure numerically, we opt for an intermediate approach: we integrate features of several proposed structures into two parametric families with a small number of parameters.  As a result,  the material property space coverage of these structures can be explored in brute-force way by parameter sweeps. This achieves two goals: first, this allows us to find an inner bound on the G-closure domain (with Hashin-Shtrikman bounds providing the outer domain). Second, we can straightforwardly tabulate the necessary structures parameters for any desirable effective elastic properties.

 \section{Periodic structure families}
 \label{sec:micro}

 Our families are based on synthesis of hexagonal/triangular isotropic structures invented by Sigmund \cite{Sigmund2000} (Figure~\ref{COMP}(C)), and the chiral structures proposed in \cite{Lakes1996,Advanced2011,Solidi2015} (Figure~\ref{COMP}(H,I)) for negative Poisson's ratio. 

\begin{figure} 
	\includegraphics[width=16cm]{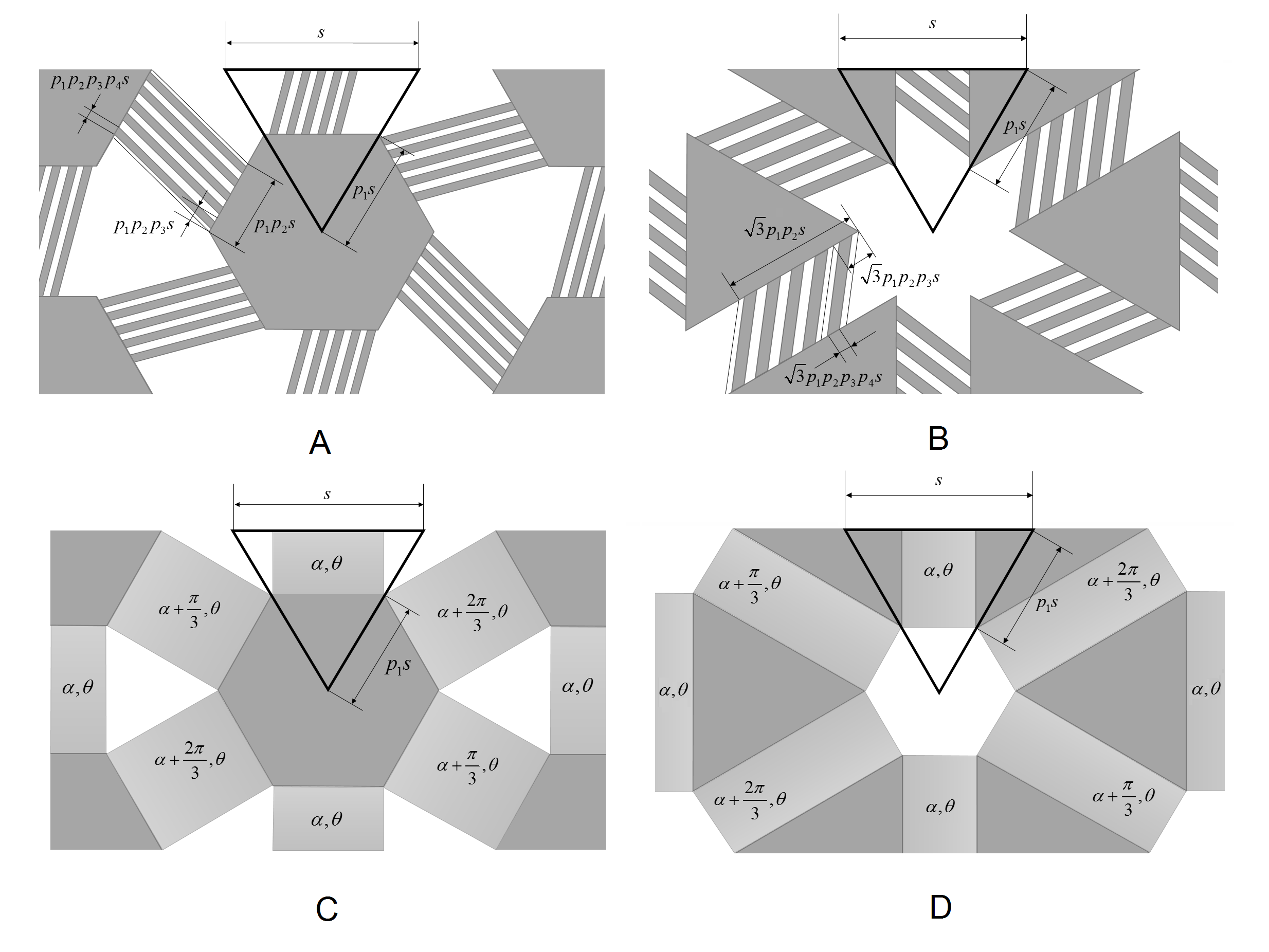}
	\caption{  (A,B) Structure families: (A) Triangle, (B) Hexagon. (C,D) Limit  structures: (C) Triangle, (D) Hexagon }
	\label{STUCT}
\end{figure}

 Figure~\ref{STUCT}(A,B) shows the proposed new periodic structures.  The starting point is the Triangle and Hexagon  structures presented in \cite{Sigmund2000}. These structures consist of triangular or hexagonal solid areas, connected by rectangles partitioned into separate beams (we refer to these rectangles as laminate areas). We extend these structures in a very simple way, by allowing the beams of laminate region to have arbitrary orientation -- this considerably extends the range of materials represented by the structure,  providing auxetic behavior via the mechanism described in \cite{Lakes1996}. Structures of this type are described by four parameters shown in the figure (size of the solid region $p_1$, width of the laminate area $p_2$, period of the beams in the laminate area $p_3$, and width of a beam $p_4$, with each dimension measured relative to the previous (\textit{e.g.},  $p_2$ is a fraction of $p_1$, in the range $0\ldots 1$). It is important to note that all parameters are \emph{continuous}, including the one determining the number of beams $p_3$; if it is not of the form $1/n$, where $n$ is an integer, the resulting structure has thinner ``remainder'' beams, which we place on the sides of the laminate area. In this way, all parameters can be changed continuously, and optimized more easily. 

We also consider a limit case of these structures, to provide a validation of the hypothesis that with the number of beams in the laminate areas going to infinity  (\textit{i.e.}, the resulting structure becomes two-scale), the material properties approach the boundary of the reachable domain (Figure~\ref{STUCT}(C)); 
these structures are parametrized in a similar way, but have only three parameters, as the beam period approaches zero. Parameter $p_1$ is defined in the same way as for finite structures, and the two remaining parameters  $\theta$ and $\alpha$ (shown in Figure~\ref{CELL}(B) below),  define the volume fraction and the orientation angle of the laminated areas.  Parameters $\alpha$ and $\theta$ are used to set the elastic tensor in the laminated region.

\section{Methods}
\label{sec:methods}

In order to perform the parametric studies of the new periodic structures, we use a FEM-based, regular-grid homogenization code, extending \cite{Xia2015}. Our algorithm of periodic homogenization follows the method of mutual energies, as described in \cite{Xia2015}.  The homogenized tensor of elasticity of the rectangular cell $Y \in \mathbb{R}^2$ is found as: 

\begin{equation} \label {hom1}
C_{ijkl}^{H}=\frac{1}{\left|Y\right|}\intop_{Y}C_{pqrs}(y)\varepsilon_{pq}^{A_{ij}}(y)\varepsilon_{rs}^{A_{kl}}(y)dY.
\end{equation}

Here $\left|Y\right|$ is the area of the cell, $C_{pqrs}(y)$ is the tensor of elasticity of the material at the point $y$ inside the cell domain, and $\varepsilon_{pq}^{A_{ij}}(y)$ are the strain fields induced by the imposition of three constant macroscopic unit test strains $A_{ij}$. These strains are found by solving three ($pq=11,22,12$) cell problems:
	
\begin{equation} \label {hom2}
\begin{split}
\left(C_{ijkl}(y) \varepsilon_{kl}^{A_{pq}}(y)\right)_{,j}=0, \\
\varepsilon_{kl}^{A_{pq}}(y)= w_{kl}(y) + A_{kl}, \\
w_{kl}(y)  \text{is Y-periodic}.
\end{split}
\end{equation}	 

\begin{figure}  
	\includegraphics[width=16cm]{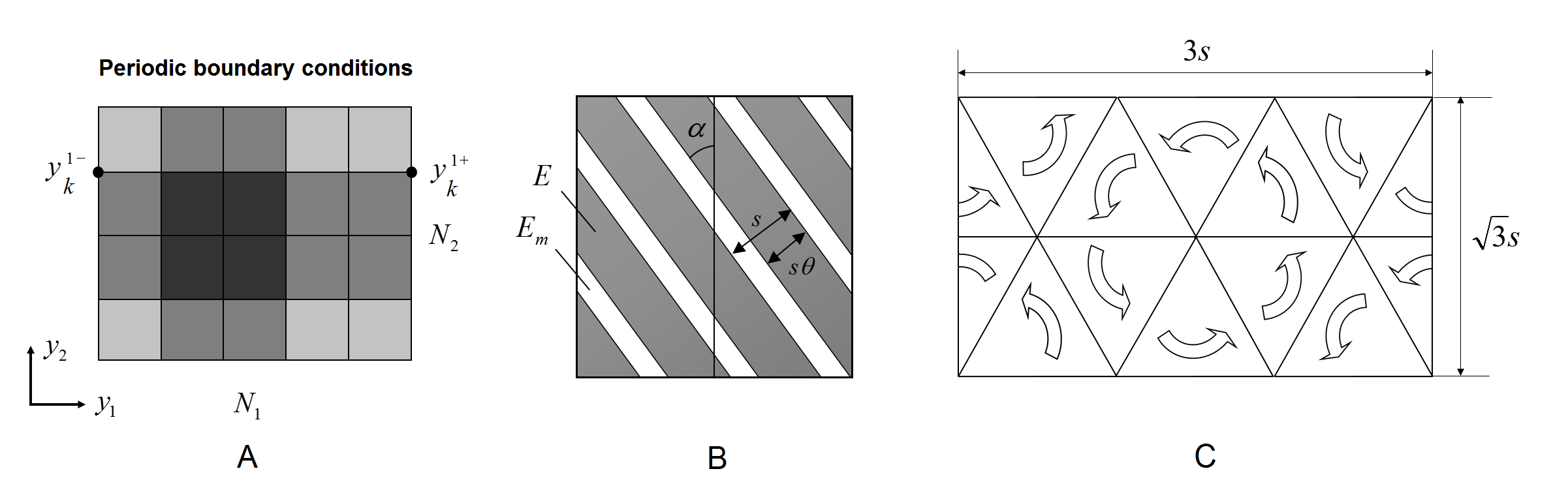}
        \caption{Diagrams  of (A) a cell computational domain (B) rank-1 laminate (C) prescribed rotational symmetries of a computational domain. }
\label{CELL}
\end{figure}

Constant unit test strains $A_{pq}$ are imposed on the domain boundary in the form of periodic boundary conditions on displacements $u_{i}^{k+}$, $u_{i}^{k-}$ in the elements nodes (Figure~\ref{CELL}(A)):

 \begin{equation} \label {pbc}
u_{i}^{k+}-u_{i}^{k-}=A_{ij}(y_{j}^{k+}-y_{j}^{k-})=A_{ij}\triangle y_{j}^{k}.
 \end{equation}
Here $k=1,2$ stands for the domain boundaries perpendicular to the $k$-th coordinate axis. The homogenized elasticity tensor is found as follows:

\begin{equation} \label {hom3}
C_{ijkl}^{H}=\frac{1}{\left|Y\right|}\sum_{e=1}^{N} {\bf u}_{e}^{A_{ij}} {\bf k}_{e}( {\bf u}_{e}^{A_{kl}}),
\end{equation}
where index $e$ stands for the element, ${\bf k}_{e}$ is the element stiffness matrix, and ${\bf u}_{e}^{A_{ij}}$ is the vector of displacements at the nodes of $e$-th element upon imposition of $A_{ij}$-th test strain.

The computational domain is the rectangle comprising $N = N_1 \times N_2$ 4-node isoparametric finite elements (Figure~\ref{CELL}(A)). For every element, we specify its density $\rho$ and the type and parameters of its stiffness matrix ${\bf k}_e$.  Finite element stiffness matrix ${\bf k}_e(D)$ is computed based on the material elasticity tensor in matrix notation $D$, which is the elasticity tensor.

For finite structures, we use isotropic elasticity tensor. 

 \begin{equation} \label {iso}
 D(E,\nu)=\frac{\rho}{1-\nu^2}  \left(\begin{array}{ccc}
 E & \nu E & 0\\
 \nu E & E & 0\\
 0 & 0 & \frac{E}{2}(1-\nu)
 \end{array}\right), 
 \end{equation}
 where $E$ and $\nu$ are the Young's modulus of either the strong or weak material (the void is approximated with a
 very weak material).
 
 To explore the limit case as the genus of the structure goes to infinity, as we increase the number of
parallel beams in a particular region, we use the anisotropic rank-one laminate tensor \cite{Allaire2002}.
 
 \begin{equation} \label{lam}
\begin{split}
  D(\alpha,\theta,E_m, E)&=\rho T(\alpha)L(\theta,E_m,E)T^{T}(\alpha),\quad \mbox{where}\\  
L(\theta,E_m,E)&=\left(\begin{array}{ccc}
(1-\theta)E_m+\theta E & 0 & 0\\
0 & \frac{1}{\frac{\theta}{E_m}+  \frac{1-\theta}{E}} & 0\\
0 & 0 & \frac{1}{2(\frac{\theta}{E_m}+\frac{1-\theta}{E})}
\end{array}\right), \\
T(\alpha)&=\left(\begin{array}{ccc}
\cos^{2}\alpha & \sin^{2}\alpha & -2 \sin \alpha \cos \alpha\\
\sin^{2} \alpha & \cos^{2}\alpha & 2\sin \alpha \cos \alpha\\
\sin \alpha \cos \alpha & -\sin \alpha \cos \alpha & \cos^{2} \alpha - \sin^{2}\alpha
\end{array}\right).
\end{split}
 \end{equation}
In this equation,   $\alpha$ is a lamination direction,  $E_m$ is the Young's modulus of the laminate's soft material (in void-material simulations $E_m=10^{-6} \cdot E$), $\theta$ is the volume fraction of the laminate's strong phase(Figure~\ref{CELL}(B)).

We ensure that the effective elasticity tensor obtained by homogenization is isotropic, by using hexagonal/triangular cells, and imposing symmetries of structures with respect to $\pi/3$ rotations, as  any elasticity tensor invariant with respect to these symmetries is isotropic \cite{Love1944}. It is important to note that reflectional symmetries are \emph{not} required for isotropy. This makes it possible for us to use \emph{chiral} structures to create isotropic materials.

We use a rectangular computational domain with the ratio of the sides $3 : \sqrt{3}$ (Figure~\ref{CELL}(C)). This rectangular base cell has the area of two hexagonal base cells.

Each square element overlapping the structure is assigned elasticity tensor of the base material (or laminate, if this part is marked as laminate area). Our drawing system uses geometric primitives (polygons) to define the structure that are rendered with anti-aliasing to obtain gray-scale density values in 0 to 1 range on a regular grid, that are used to scale the elasticity tensor.

 The net effect of replacing the 
	precisely defined structure with its  anti-aliased rasterization is hard to estimate precisely, although it can be expected to improve accuracy vs. using a binary density. Experimentally, we have observed that this discretized family of structures has Poisson ratio range for a given Young modulus, to the family of  obtained by conforming triangulation of the precise structure, as discussed in more detail in Appendix A. The regular grid approximation is significantly faster to evaluate.

For the verification purposes, our computations were also checked with an alternative homogenization tool \cite{Zorin2015}, which is based on representing the shape as an implicit function, and meshing the domains with an unstructured triangular mesh.  These two approaches yield close results -- see Appendix A for  details.  

Our homogenization and structure generation codes are available at \url{https://bitbucket.org/iostanin/pco_toolbox_matlab}.

\section{Numerical studies}
\label{sec:results}

We explore the coverage of the families of structures described above using homogenization under several scenarios, as well as provide evidence, using shape optimization, that the boundary we obtain is likely to be close to optimal.  For simplicity and ease of understanding, we use a base material
with Poisson's ratio $\nu_b = 0$. However, our results can be straightforwardly generalized to any Poisson's ratio, thanks to the CLM theorem \cite{CLM_theorem}.  The specific value of $E_b$ simply sets the scale, as we use linear elasticity, so in all plots we use the ratio $E/E_b$ for the effective Young's modulus $E$.   

We define the \emph{coverage} of a family of structures as the set of points in effective material properties space that can be obtained by varying structure parameters. In order to quantify coverage, we introduce coverage coefficients $\Delta_R, \Delta_L$, defined as ratios of the highest (lowest) achievable Poisson's ratio at  $E/E_b = 0.5$, and the corresponding theoretical limit value of Poisson's ratio at  $E/E_b = 0.5, \phi = 1$ (in case of $\nu_b = 0$, these limit values are $\nu_R = 0.5$, and $\nu_L = -0.5$).  

\begin{figure}  
\includegraphics[width=16cm]{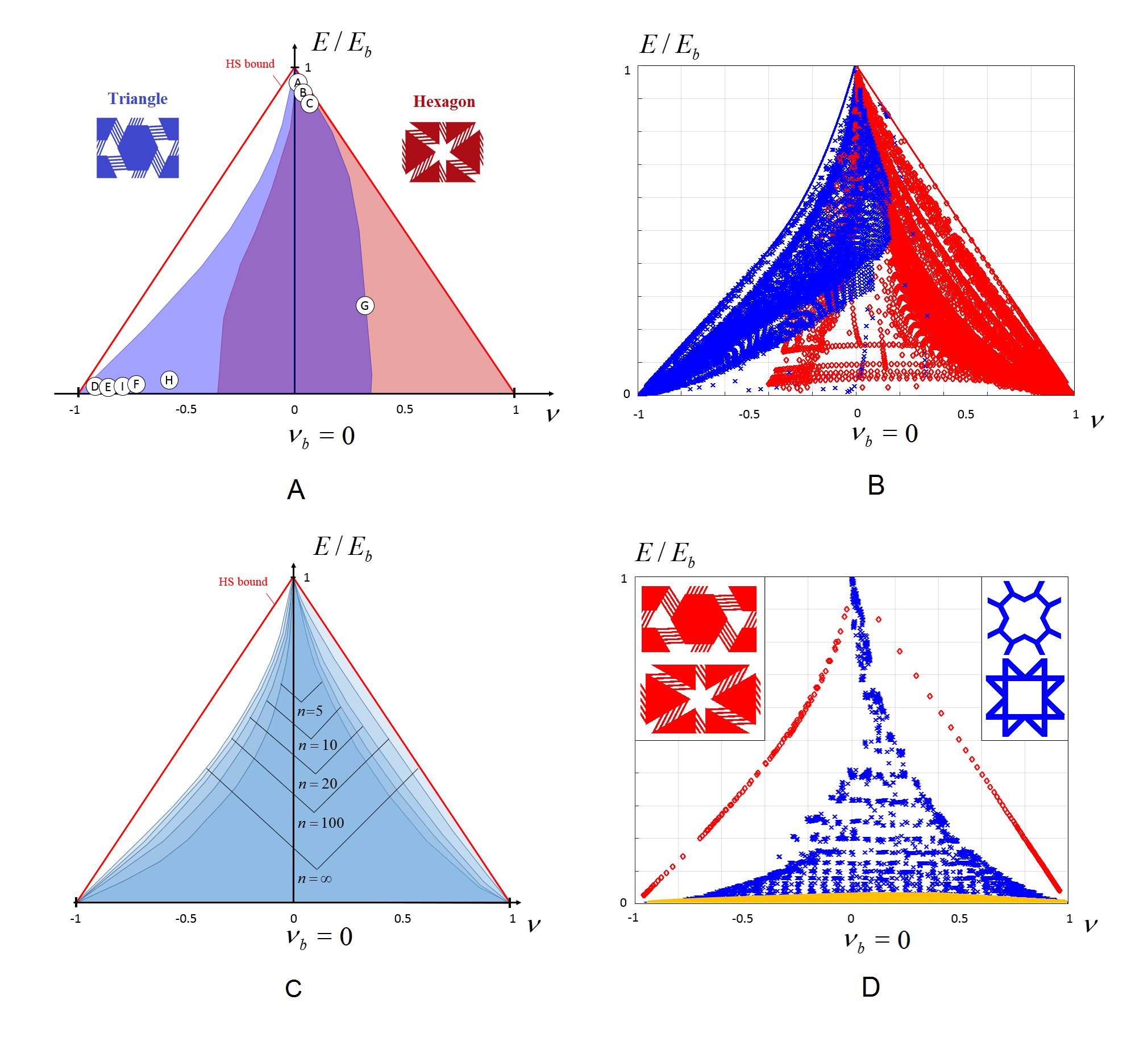}
\caption{  (A, B) Coverage of Hashin-Shtrikman bounds with the new structures of all types. Letters denote tentative positions of the extremal composites presented in Figure~\ref{COMP}. (A) - covered sets, (B) - raw data samples and the conjectured analytical bounds.  (C) Expansion of the coverage as the number of beams $n=1/p_3$ increases. (D) Comparison of the coverage obtained in this work with Triangle and Hexagon structures with ($n = 30$, $\nu_b = 0$,  only extreme structures are shown, red points), in comparison with the widest coverages obtained based on previous numerical studies \cite{Zorin2015,Zorin2017} (blue points), and analyticaly computed coverage achieved with isotropic laminates proposed in \cite{Milton1992} (yellow region, parameters of lamination are sweeped maintaining isotropy, the ratio between stiffnesses of weak and strong phases is $10^{-6}.$}  
\label{COVERAGE}
\end{figure}

Figure~\ref{COVERAGE}(A, B) shows the coverage we were able to obtain with two structure families (Triangle and Hexagon). Figure~\ref{COVERAGE}(A) provides the areas covered by two structures, whereas Figure~\ref{COVERAGE}(B) shows raw datasets, as well as the analytical conjectured bounds based on these datasets (see the discussion below). For comparison, we also sketch the approximate positions of several  known extremal composites, shown in Figure~\ref{COMP}. Figure \ref{COVERAGE}(C) gives an idea how the coverage depends on the number of beams $n = 1/p_3$ for single scale structures.

Relatively few papers attempted to design families of structures with a large coverage area. The best coverage known to us was obtained in  \cite{Zorin2015,Zorin2017} in 3D. For comparison, we applied the same approach to two-dimensional structures with square symmetries, to estimate coverage that can be compared to the families of structures considered in this paper. 

More specifically, a ground structure was obtained by subdividing a square into smaller squares and triangulating each of these.
These ground structures are parametrized by node sizes and displacements. 
Initially, a combinatorial search is performed by removing edges from the ground structure (while maintaining the square symmetries)
and generating a set of initial structures that are estimated to yield the largest range of material properties.
Then the parameters of each identified structure are optimized multiple times, to minimize the deviation of the homogenized
material properties from a set of target pairs $(E_i, \nu_i)$ in the admissible region. This optimization is performed
by solving the adjoint equations of elasticity at every step to compute the shape derivative. The topology of each
structure remains fixed during the optimization. 

Figure \ref{COVERAGE}(D) gives the comparison of the performance of our new structures with the ones found previously by ground-state combinatorial search with subsequent shape optimization which provided the widest coverage known so far. We use coverage data for two structures, shown in the same figure, that cover essentially the whole domain covered by any structure obtained by
edge removal. We can see that our new Triangle and Hexagon structures ($n = 30$) significantly improve the coverage of possible elastic properties of composites for the same base material's elastic moduli.
Fig 6(D) also provides the coverage of effective elastic properties achievable with the isotropic laminates described in \cite{Milton1992} (yellow line). As one can see, in case of vanishing weak phase and negative Poisson's ratio, these laminates appear to be much more compliant than our microgeometries.

Below we take a closer look at the obtained numerical results.

\paragraph{Optimality of Hashin-Shtrikman bounds for positive Poisson's ratios}

\begin{figure}  
	\includegraphics[width=16cm]{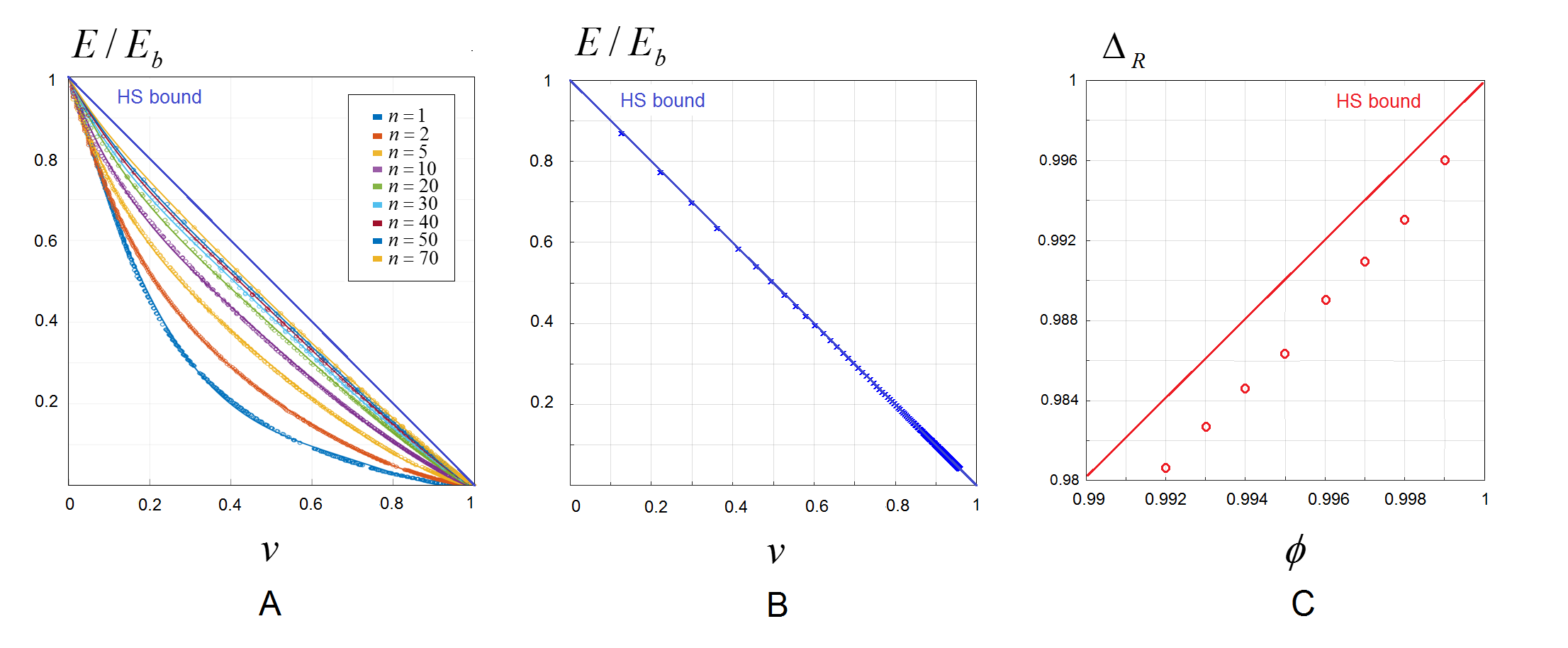}
	\caption{  (A) Coverage of elastic properties for positive Poisson's ratios as a function of the number of beams in the joint (B) Coverage achieved in the limit case of two-scale structure (C) coverage measure $\Delta_R$ as a function of volume fraction $\phi$ }
	\label{RIGHT_SIDE}
\end{figure}

Our results provide compelling numerical evidence that the right Hashin-Shtrikman bound for non-fixed volume fraction can be approached with good precision with Hexagon structures as the number of beams $n$ approaches infinity ($n = 1/p_3$). This is confirmed using simulation with limit laminate structures, which approach the boundary very closely. Figure \ref{RIGHT_SIDE}(A) demonstrates the increase of coverage with increase of $n$, figure \ref{RIGHT_SIDE}(B) demonstrates the limit case of the structures with $\phi \rightarrow 1, n \rightarrow \infty$, achieving right Hashin-Shtrikman bound. Figure \ref{RIGHT_SIDE}(C) indicates that limit laminate structures also remain very close to Hashin-Shtrikman bound for volume fractions smaller than $1$ (see also Figure \ref{volfrac}). These results are consistent with the ones obtained previously in \cite{Sigmund2000} and recent work \cite{Milton2017}, establishing the attainability of the right Hashin-Shtrikman bound for volume fractions smaller than 1.   

We observe that chirality (captured by the parameter $p_2$, or $\alpha$ for a limit two-scale structure ) does not play a major role in this case, \textit{i.e.} the right Hashin-Shtrikman bound is closely approximated by structures with no chirality.

 The stiffness of such structures (for the considered case $\nu_b=0$ -- the position $(E, v)$ on $E + \nu = 1$ straight line) is defined solely by the stripe's relative thickness ($p_4$ for type one structure and $\theta$ for the limit two-scale structure).   

\paragraph{Best achievable bound for negative Poisson's ratios}

\begin{figure}  
	\includegraphics[width=16cm]{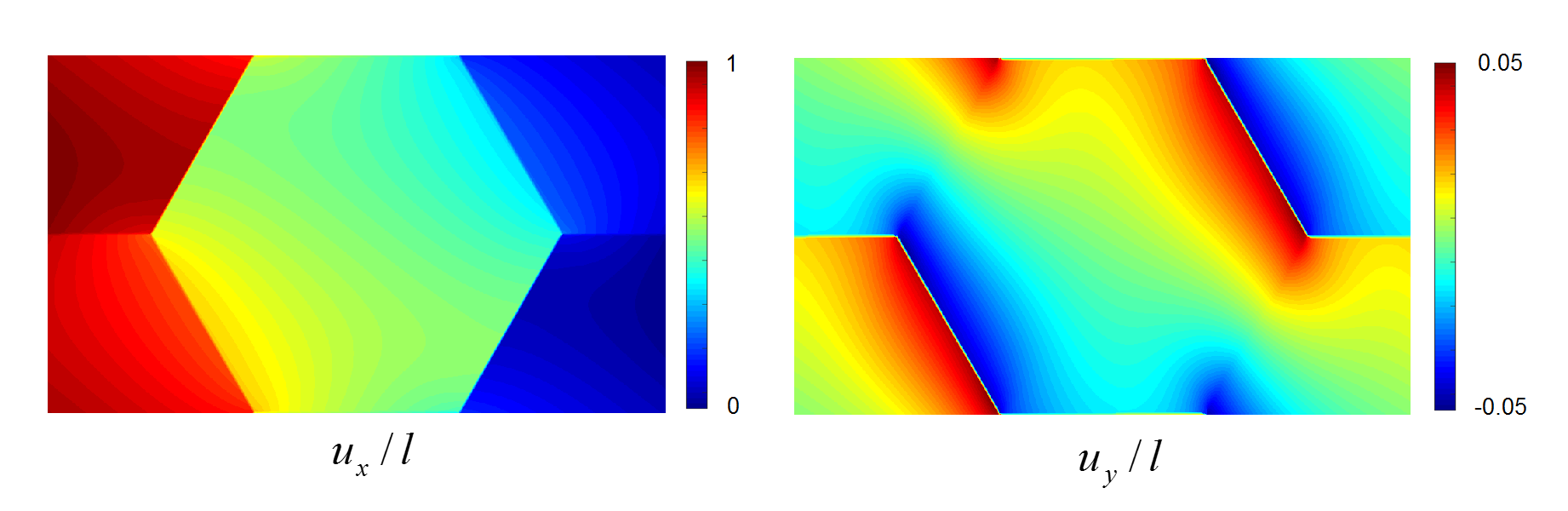}
        \caption{ Displacement fields in the horizontal tension ($pq=11$, Triangle limit structure) cell problem.
          Rotation of solid parts of the structure, leads to auxetic behavior.  }
\label{ROTATION}
\end{figure}

\begin{figure}  
	\includegraphics[width=16cm]{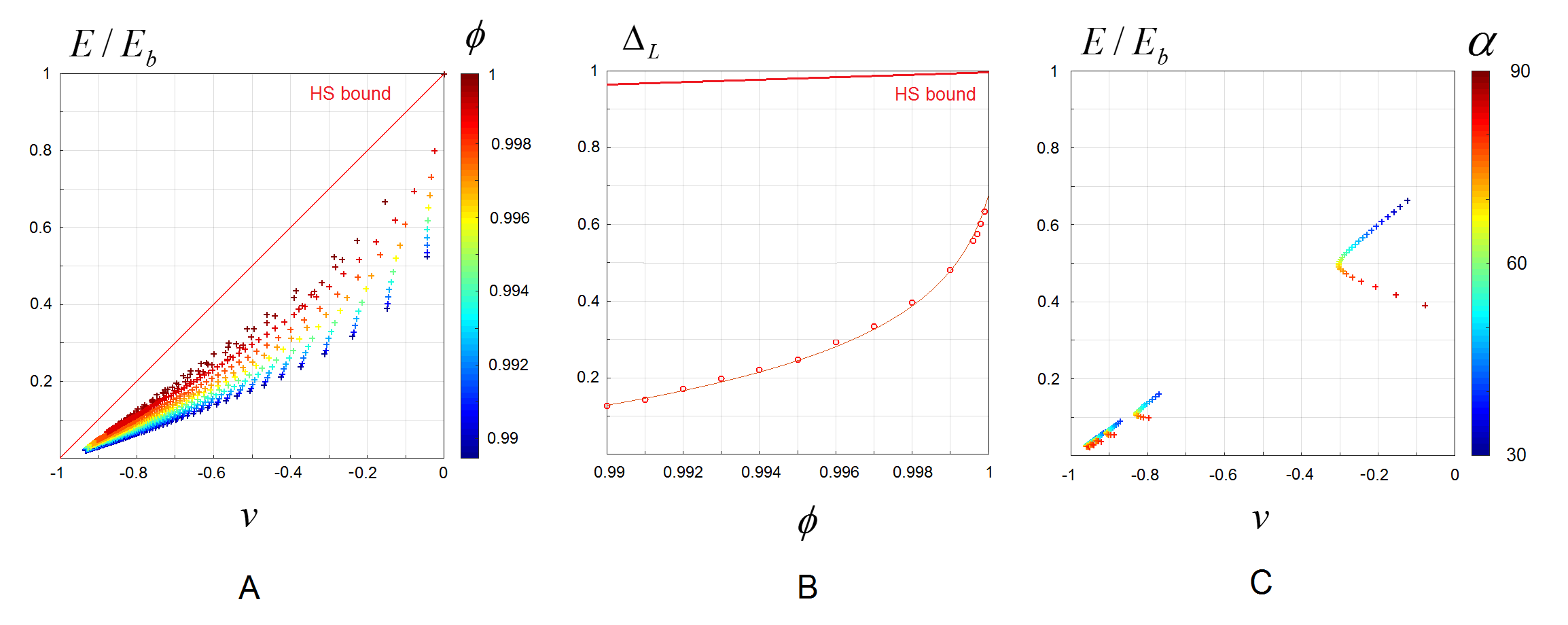}
	\caption{ Limit structures approaching left HS bound. (A,B) Left-bound extreme elastic moduli, achieved with the Hexagon limit structure; chart colored by values of $p_1$ (A) and $p_2$ (B). (C) Dependence of the achievable negative Poisson's ratio as the function of chirality angle. Chirality angle $p_3$ sweeps are performed for several values of $p_2$.}
\label{chirality}
\end{figure}

\begin{figure}  
	\includegraphics[width=16cm]{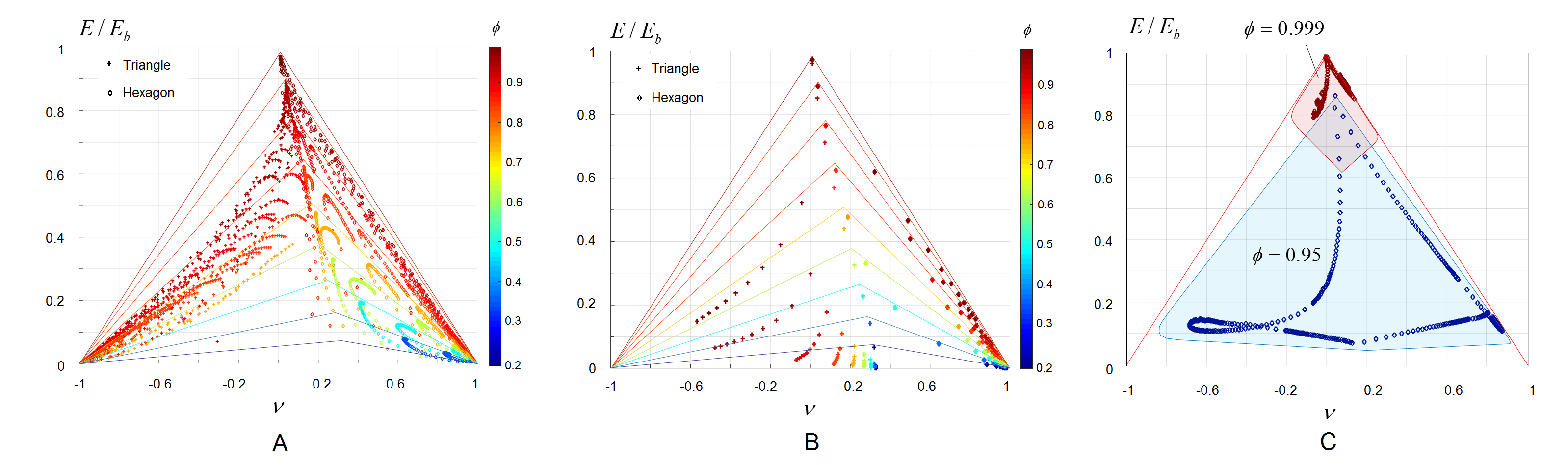}
	\caption{ Coverage of Hashin-Shtrikman bounds with triangular and hexagonal structures for a set of intermediate volume fractions. (A) Structures of the first type, (B) structures of the second type. (C) Coverage of Cherkaev-Gibiansky bounds with bimaterial composite for two volume fractions.}
	\label{volfrac}
\end{figure}

Introduction of nonzero chirality leads to a decrease of the effective Poisson's ratio due to the rotation of solid segments of the structure (Figure~\ref{ROTATION}).  Extremal values that closely approach the left theoretical bound (Figure~\ref{chirality}(A)) are achieved by the Triangle structure, which is, in a sense, a geometric dual of the Hexagon structure. Similarly to the right bound, increasing the number of beams in the laminate area leads to structures closer to the boundary (Figure \ref{COVERAGE}(C).  However, in this case, the limit structure does \emph{not} follow the boundary. Figure~\ref{chirality}(B) demonstrates this coverage coefficient as a function of the volume fraction $\phi$. The coverage curve grows rapidly when approaching volume fraction 1. However, reasonable extrapolation indicate that the coverage does not exceed 0.7. The angle of chirality $\alpha$ of the extreme structures is always close to $\pi/3$, however, it depends slightly on the Young's modulus (Figure~\ref{chirality}(C)).

The right bound achieved in our numerical tests is reasonably well approximated by a fourth order polynomial (Figure~\ref{COVERAGE}(B)):

\begin{equation}
E(\nu)=1+(1+c_{1}-c_{2}+c_{3})\nu+c_{1}\nu^{2}+c_{2}\nu^{3}+c_{3}\nu^{4}
\end{equation}

with the constants $c_{1} = 3.477$, $c_{2} = 3.098$, $c_{3} = 0.988$. 

\paragraph{Intermediate volume fractions}

Above we could see that our Triangle and Hexagon structures exhibit wide coverage of elastic properties for volume fractions approaching $1$. Next, we examine the performance of these structures for intermediate volume fractions. 

For both Triangle and Hexagon structures the volume fraction is given by:
\begin{equation} \label {vf}
\phi = 2 p_4 p_1 (1 - p_1) + p_1^2, 
\end{equation}
or, for the limit two-scale structure,
\begin{equation} \label {vf2}
\phi = 2 \theta p_1 (1 - p_1) + p_1^2. 
\end{equation}
Considering that for the extremal composites $p_4 (\theta)$ has to stay very close to $1$, the volume fraction can be viewed primarily as the function of a single argument $p_1$.

Figure~\ref{volfrac} illustrates the coverage of Hashin-Shtrikman bounds for a set of intermediate volume fractions. Figure~\ref{volfrac} (A) shows the parametric sweeps for structures with fixed number of beams, Figure~\ref{volfrac} (B) demonstrates the behavior of their two-scale counterparts. 

For both single scale and two-scale structures, we explored the parameter $p_1$ in the range $0.1 \ldots  0.9$. We can see that in both cases the right Hashin-Shtrikman bound is easily approached with Hexagon structures with zero chirality angle, however, the coverage on the left side deteriorates dramatically with decrease of the base material volume fraction. This effect has trivial explanation. Consider the case of two-scale limit structure. Clearly, if $\theta < 1$ and the lamination angle is steeper than the diagonal of the rectangular joint region, the stiffness of the joint drops to zero. Therefore, for the values of $p_1$ smaller than the critical value defined by 
 
 \begin{equation} \label {ang}
 \sin{\alpha} = \frac{p_1}{\sqrt{ 3 (1-p_1)^2  + p_1^2 }},
 \end{equation}
the stiffness of the Hexagon structure will drop to zero. For $\alpha = \pi/6$, the critical value of $p_1 = 0.5 $, and the corresponding volume fraction $\phi = 0.75$.  Figure~\ref{volfrac}(B) shows that all  structures with this volume fraction and  $p_2<1$ have nearly zero stiffness, with the remaining response is determined by the nonzero density of the ersatz material.

\paragraph{Non-void weak material}
In case of a bimaterial mixture, the range of the elastic properties that could be achieved with our structures shrinks significantly compared to the corresponding theoretical bounds. Figure~\ref{volfrac}(C) gives the coverage achieved for the composite material with the weak phase $1000$ times softer than the strong phase. For these cases we performed same parameter sweeps that provided us the most extreme properties in the case of void-material composite. We can see that though the structures without chirality remain close to right bound, the chiral bimaterial structures are very far from the extreme properties.

\paragraph{Expanding the parameter space} 

\begin{figure}  
	\includegraphics[width=16cm]{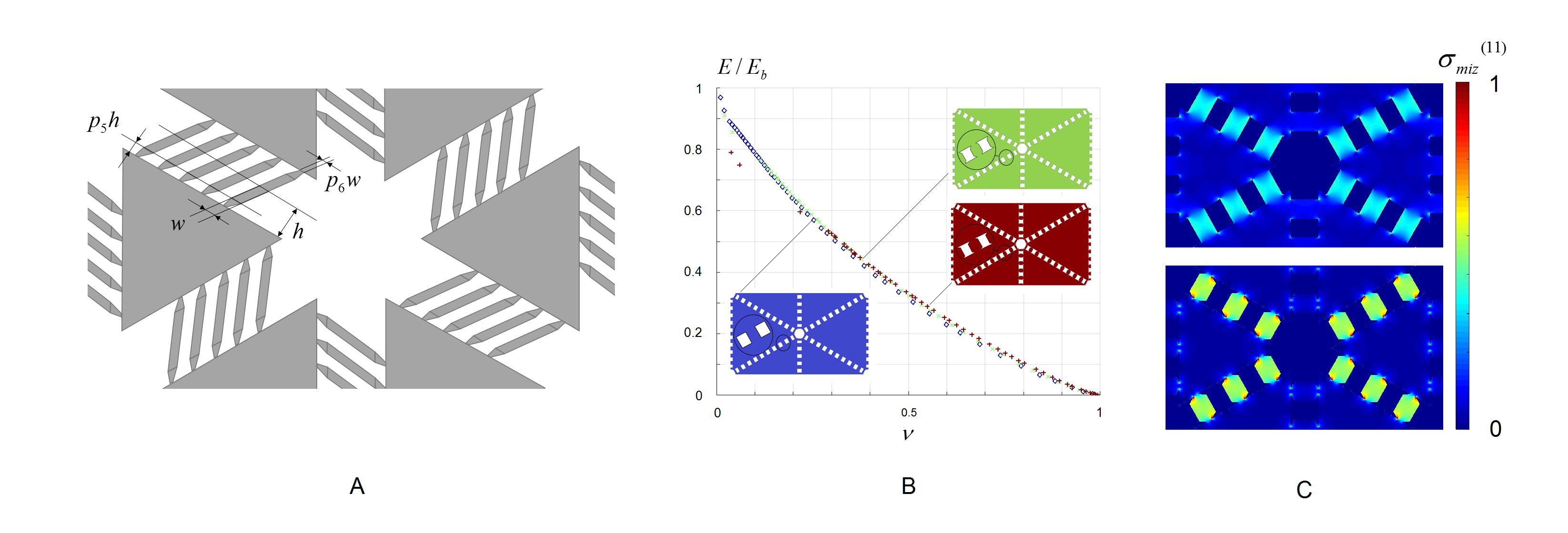}
	(A)  Expanding parameter space: a parameterization of the structure with indented beams. 
	\caption{(B) Change in coverage with addition of indentations. (C) Von Mises stress concentrations caused by the introduction of beam indentations. }
	\label{improve-coverage}
\end{figure}

Potentially, expanding  the parameter space may improve the performance of our structures. One obvious way to improve the performance is to modify the configuration of joints, that are supposed to be very strong in axial compression, but very compliant in shear. The performance of the joints can be improved by introducing "indentation" regions at the tips of the beams (Figure \ref{improve-coverage}(A)). In our numerical experiments the structures with indented beams does not improve the coverage of the elastic moduli at large volume fractions, since the features of sharpened stripes simply can not be resolved. However, they can be useful for the case of intermediate volume fractions -- low values of shear modulus can be reached at a lower complexity of the structure (\textit{i.e.} smaller number of stripes), which can be important in practical situations. Figure~\ref{improve-coverage}(B) illustrates the improvement of the coverage achieved with the structures with intended beams compared to the original ones. However, this relatively insignificant improvement comes at a cost -- higher stress concentrations at the tips of the stripes, that make failures more likely (Figure~\ref{improve-coverage}(C)). 

\paragraph{Optimality of the negative Poisson's ratio bounds}  

It is difficult to establish with certainty that no points closer to the Hashin-Shtrikman bounds can be found. However, it is highly likely that a very different structure would be needed for this. To validate this conjecture numerically, we have performed the following numerical experiments. Starting from a structure with effective elastic parameters far away from the achievable boundary (but of sufficiently large number of beams), we vary the parameters of the structure to optimize the goal functional.   We consider the functional $J = \frac{1}{2} \|S^H - S^*\|^2_F$, the Frobenius norm of the difference between the effective elasticity tensor of our structure, and a target tensor $S^*$, which we take to be a point on the Hashin-Shtrikman boundary $(-0.5,0.5)$. We optimize this functional until the point reaches a minimum which is on the achievable boundary. 

At this point, the gradient of the functional, with respect to the structure parameters, is zero (or, rather,
its projection on the hyperplane orthogonal to the constraint gradients). This of course, does not imply optimality,
as we are considering only a small number of parameters. However,  we can measure the norm of the shape derivatives vector 
with respect to \emph{positions of all points on the shape boundary}.

The shape derivative measures how perturbing the shape affects the objective function 
(given by $J = \frac{1}{2} \|S^H - S^*\|^2_F$). To compute it, we use the following formulation 
\cite{Zorin2015}:
\begin{align*}
  dJ[v]   &= [S^H - S^*] : dS^H[v]\\
  dS^H[v] &= -S^H : dC^H[v] : S^H\\
  dC^H[v] &= \frac{1}{|Y|} \int_{\partial w} [(A_{ij} + \epsilon(w_{ij})):C^{\text{base}} : (A_{kl} + \epsilon(w_{kl}))]\,  (v \cdot \hat n)\, dA(y)
\end{align*}
Here, $v$ is a vector field corresponding to the velocity of the shape. Also, as described before,
$A_{kl}$ represents the unit test strain, while $w_{kl}$ is the microscopic fluctuation in base cell $Y$.

In the discrete setting, the shape derivative can also be written as a sum of inner products: $\sum_m sd_m \cdot v_m$, where $v_m$ is the velocity on vertex $m$ of the input mesh and $sd_m$ represents the shape derivative at this same node. This enable us to compute the norm of the derivative.

 The values of shape derivative are tabulated in Table 1, as a function of the iteration number. We see that close to the boundary the derivative declines by more than an order of magnitude. Examining the distribution of nonzero values, we see that these are almost
entirely close to sharp corners of the structure, and are close to zero elsewhere. This suggests that adding degrees
of freedom to optimization is not likely to improve how close one can get to the Hashin-Shtrikman bound for negative
Poisson's ratios. The bound we observe appears to be very close to a local optimum for the topology we consider.  This does not imply golbal optimality of our structures; however, a large change in geometry or topology is likely to be  needed for further improvement.

\begin{table}
  \label{tab:shapederiv}
  \caption{Norm of the shape derivatives vector as a function of iteration number}
 \begin{tabular}{|c|c|c|c|}
 	\hline 
 	N & $\left\Vert \nabla\right\Vert_2 $ & N & $\left\Vert \nabla\right\Vert_2 $\tabularnewline
 	\hline 
 	\hline 
 	1 & 172.64 & 11 & 19.24\tabularnewline
 	\hline 
 	2 & 170.82 & 12 & 18.65\tabularnewline
 	\hline 
 	3 & 169.76 & 13 & 13.73\tabularnewline
 	\hline 
 	4 & 168.05 & 14 & 13.67\tabularnewline
 	\hline 
 	5 & 165.05 & 15 & 10.15\tabularnewline
 	\hline 
 	6 & 117.15 & 16 & 10.37\tabularnewline
 	\hline 
 	7 & 90.77 & 17 & 8.78\tabularnewline
 	\hline 
 	8 & 90.54 & 18 & 9.00\tabularnewline
 	\hline 
 	9 & 51.56 & 19 & 9.03\tabularnewline
 	\hline 
 	10 & 31.42 & 20 & 8.67\tabularnewline
 	\hline 
 \end{tabular}

 \end{table}

\section{Discussion, conclusions and future work}
\label{sec:conclusions}

In our work we have demonstrated nearly complete G-closure for positive Poisson's ratio -- effectively, attainability of Hashin-Shtrikman bounds for the elastic properties. While for negative Poisson's ratio we have not reached the bound, and it remains unclear whether Hashin-Shtrikman bounds are  optimal in this case,
presented family of structure has broader coverage that any previously considered family. 

The region that is left not covered by our structure is the neighborhood of the left HS bound with values of Young's modulus close to 0.5. A number of alternative approaches (\textit{e.g} topology optimization) did not allow us to reach the extremal composite that would attain an isotropic elastic tensor in that region. The questions of the existence of such microstructures, as well as the optimality of Hashin-Shtrikman bounds in that neighborhood remains unanswered. 

The range of achievable elastic properties available with our two microstructures covers all known two-dimensional isotropic void-material composites (some of them are shown in Figures 3 and 6(A,D)). We are not aware of the regular isotropic structures, random assemblages or general topology optimization solutions that provide elastic moduli that fall out of the coverage presented in our work. Therefore, our work suggest the most robust solution of the two-dimensional problem of inverse homogenization of void material isotropic composites available today.

\begin{figure}  
  \includegraphics[width=16cm]{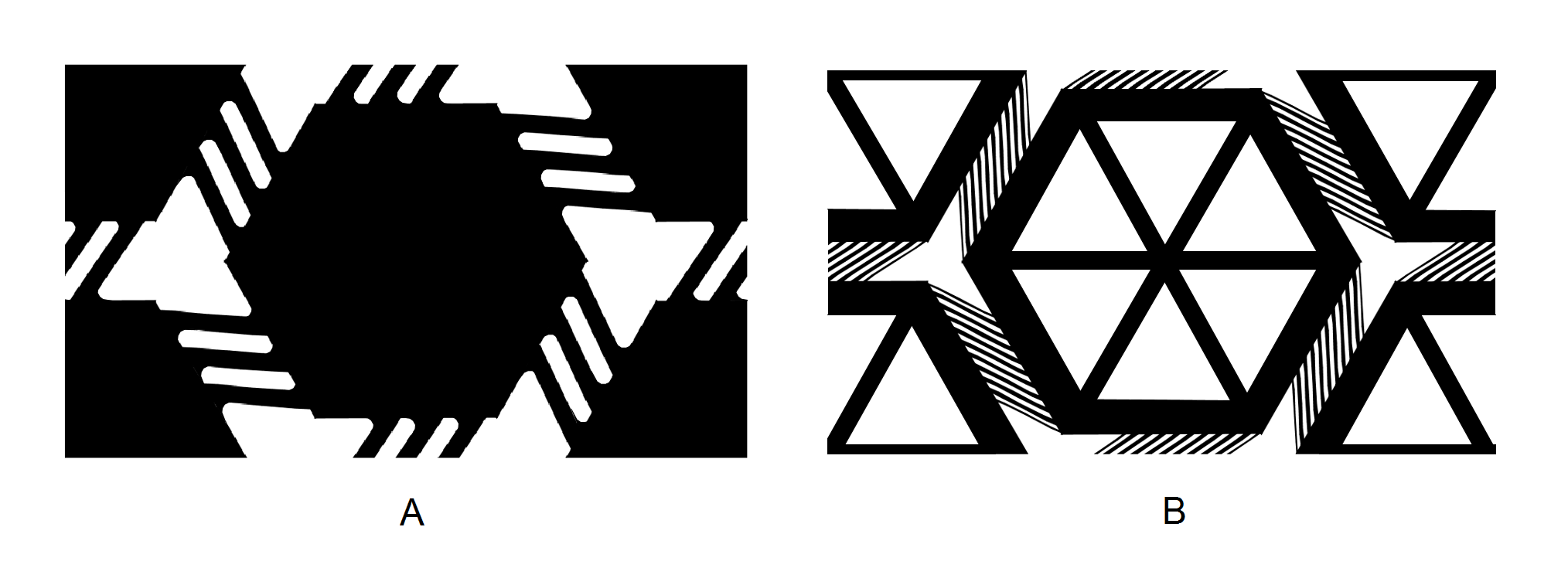}

  \caption{Possible generalizations of the extremal structures.
    (A) Structure with an improved strength due to elimination of stress concentrations (B) Structure with wide coverage of elastic moduli at intermediate volume fractions. }
\label{generalization}
\end{figure}

The theoretical structures suggested in our work can be adapted for practical usage. For example, sharp corners causing stress singularities can be smoothened out (Figure~\ref{generalization}(A). On the other hand, a straightforward modification of the structure (Figure~\ref{generalization}(B)) can dramatically improve the performance of the microstructures with intermediate volume fractions.

Although the microstructures suggested in this work do not allow straightforward generalization of the three-dimensional case, similar ideas (laminate-like regions and chirality) can be used to improve the performance of the earlier developed 3D patterns with programmable elastic properties \cite{Zorin2015,Zorin2017}. 

\section{Acknowledgments} 

The work was supported by the Russian Science Foundation through grant RSF-15-1100033. I.Ostanin acknowledges the financial support from the Russian Foundation of Basic Research (RFBR) under grant 16-31-60100.

\section{References}

\bibliographystyle{natbib}
\bibliography{manuscript}

\section*{Appendix A - Approximation accuracy}

The discretization used in our study introduces errors in the behavior of homogenized elastic constants, 
we have compared our approach to a more precise, but significantly slower approach based on unstructured 
conforming meshing of structure boundaries. 

The  counteracting sources of this discrepancy is inadequate stiffness of inclined beam elements that are represented on a regular grid and replacement of very low stiffness soft phase with a stiffer material. 
To compare effective properties achieved by the two discretization, we match the effective Young modulus 
for two structures, by varying the beam thickness ($p_4$) while keeping other parameters the same 
and compare the effective Poisson ratio. One achieves the accuracy better than one percent (Table~\ref{tab:compare2}) on several samples for both structures,  which indicates that the coverage obtained by using the alternative slow method is very close.

This level of accuracy applies only for the situations when the grid refinement is sufficient to resolve fine features, beam elements and notches between them. 

\begin{table}
	\label{tab:compare2}
	\caption{Matching regular grid computations and unstructured mesh computations. Beam thicknesses parameter for the regular grid is chosen to match the values of the Young's modulus found on an unstructured grid}
\begin{tabular}{|c|c|c|c|c|c|c|c|c|}
	\hline 
	Type & Grid & $E$  & $\nu$ & $p_{1}$ & $p_{2}$ & $p_{3}$ & $p_{4}$ & $\varepsilon$\tabularnewline
	\hline
	\hline 
	\multirow{6}{*}{Hexagon} & Reg. & 0.3553 & 0.4917 & 0.75 & 0.1 & 1.0 & 0.77 & \multirow{2}{*}{0.0073}\tabularnewline
	\cline{2-8} 
	& Unstr. & 0.3554 & 0.4873 & 0.75 & 0.1 & 1.0 & 0.74 & \tabularnewline
	\cline{2-9} 
	& Reg & 0.3121 & 0.4595 & 0.74 & 0.2 & 1.0 & 0.546 & \multirow{2}{*}{0.0173}\tabularnewline
	\cline{2-8} 
	& Unstr. & 0.3123 & 0.4500 & 0.74 & 0.2 & 1.0 & 0.526 & \tabularnewline
	\cline{2-9} 
	& Reg & 0.6846 & 0.1734 & 0.88 & 0.1 & 1.0 & 0.657 & \multirow{2}{*}{0.0111}\tabularnewline
	\cline{2-8} 
	& Unstr. & 0.6845 & 0.1656 & 0.88 & 0.1 & 1.0 & 0.642 & \tabularnewline
	\hline 
	\hline 
	\multirow{6}{*}{Triangle} & Reg. & 0.1408 & -0.5154 & 0.73 & 0.2 & 0.7 & 0.805 & \multirow{2}{*}{0.0098}\tabularnewline
	\cline{2-8} 
	& Unstr. & 0.1409 & -0.5205 & 0.73 & 0.2 & 0.7 & 0.787 & \tabularnewline
	\cline{2-9} 
	& Reg & 0.0760 & -0.7337 & 0.84 & 0.2 & 0.65 & 0.794 & \multirow{2}{*}{0.0051}\tabularnewline
	\cline{2-8} 
	& Unstr. & 0.0759 & -0.7375 & 0.84 & 0.2 & 0.65 & 0.769 & \tabularnewline
	\cline{2-9} 
	& Reg & 0.7685 & 0.0112 & 0.93 & 0.2 & 0.9 & 0.846 & \multirow{2}{*}{0.0143}\tabularnewline
	\cline{2-8} 
	& Unstr. & 0.7681 & 0.0010 & 0.93 & 0.2 & 0.9 & 0.812 & \tabularnewline
	\hline 
\end{tabular}
\end{table}
\end{document}